\documentstyle[12pt]{article}

\begin{document}

\begin{flushright}
IMSc/2001/01/06 \\
hep-ph/0101242 
\end{flushright}

\vspace{2ex}

\begin{center}

{\large \bf  Some Phenomenological Aspects of the  } \\ 

\vspace{2ex}

{\large \bf $(n + m + 1)$ dimensional Brane World } \\

\vspace{2ex}

{\large \bf  Scenario with an $m$-form Field } \\

\vspace{8ex}

{\large  S. Kalyana Rama}

\vspace{3ex}

Institute of Mathematical Sciences, C. I. T. Campus, 

Taramani, CHENNAI 600 113, India. 
 
\vspace{1ex}

email: krama@imsc.ernet.in \\ 

\end{center}

\vspace{6ex}

\centerline{ABSTRACT}
\begin{quote} 
In the $D = (n + m + 1)$ dimensional brane world scenario with
$m$ compact dimensions, the radion modulus can be stabilised by
a massive bulk $m$-form antisymmetric field. We analyse some of
the phenomenological aspects of this scenario. We find that the
radion mass is smaller than the TeV scale, but larger than that
in the case where the radion modulus is stabilised by a bulk
scalar field. From the macroscopic $n$ dimensional spacetime
point of view, the $m$-form field mimics a set of $p$-form
fields. We analyse the mass spectrum of these fields. The lowest
mass is $\stackrel{>}{_\sim} TeV$ whereas, for any bulk or brane
field, the excitations in the compact space have Planckian mass
and are likely to reintroduce the hierarchy problem. Also, we
analyse the couplings of the $m$-form field to the matter fields
living on a brane. The present results are applicable to more
general cases also.  
\end{quote}

\vspace{2ex}

PACS numbers: 11.10.Kk, 04.50.+h, 11.25.Mj 

\newpage

\vspace{4ex}

{\bf 1.}  
Randall and Sundrum have recently proposed a simple five
dimensional brane world scenario to solve the hierarchy problem
\cite{rs1}.  Here, the extra spatial dimension has the topology
${\bf S^1/Z_2}$ in which two branes are located at the fixed
points of the ${\bf Z_2}$ with one of them representing our
universe.  For suitable bulk and brane potentials, the spacetime
metric becomes warped which essentially solves the hierarchy
problem.  See \cite{rs1} for details.

However, it is crucial to stabilise the size of ${\bf S^1}$, the
radion modulus, at the required value with no fine tuning.
Goldberger and Wise have shown \cite{gw} that the radion modulus
can indeed be stabilised by a massive bulk scalar field. 
See \cite{gw,gubser} for details. (For radion stabilisation through 
quantum effects at non zero temperature, see \cite{odin}.)

Various phenomenological aspects of this scenario have been
analysed. In particular, the radion mass spectrum have been
analysed both perturbatively, in the background of the warped
metric \cite{randall,gw2}, and exactly, including the back
reaction of the scalar field on the warped metric
\cite{gubser,tanaka,csaki}. Fluctuation spectrum of a bulk
scalar field \cite{gw0} and a vector field
\cite{hewett,pomarol}, as well as their couplings to matter
fields living on a brane, have also been analysed in the
background of the warped metric.

Alternatively, the radion modulus can also be stabilised
\cite{k} by a massive bulk $m$-form antisymmetric field
\footnote{The $m$-form fields have also been considered in other
contexts \cite{pform}.} in $D = (n + m + 1)$ dimensional
spacetime with $m$ compact dimensions, with $n = 4$
corresponding to the observable case. Such massive $m$-form
fields appear naturally in the ten dimensional massive type IIA
supergravity \cite{romans}, where $m = 2$, or in the `new
massive type IIA supergravity' \cite{howe}, where $m = 1, 3$.
Although it is not clear whether the present scenario can be
realised in such fundamental theories, it is nevertheless
important to analyse the phenomenological aspects, in particular
those that can distinguish the present scenario from that of
\cite{rs1,gw}.

In this letter, we analyse these aspects. We analyse the radion
mass spectrum in the background of the warped metric \footnote{
A rigorous analysis, as in \cite{tanaka,csaki}, requires an
exact solution including the back reaction of $m$-form field on
the metric which, however, is not known at present. Hence, we
analyse the spectrum perturbatively as in \cite{randall,gw2}.}.
Note that from the macrosocopic $n$ dimensional spacetime point
of view, the $m$-form field mimics a set of $p$-form fields. We
analyse the mass spectrum of these fields. Also, we analyse the
couplings of the $m$-form field to the matter fields living on a
brane.

The main distinguishing features we find are the following.  We
find that the radion mass is smaller than the TeV scale, but
larger than that in \cite{randall,gw2,tanaka,csaki}.
Furthermore, there will be $\frac{m!}{p! (m - p)!}$ number of
$p$-form fields, $1 \le p \le min (n, m)$, in the macroscopic
$n$ dimensional spacetime.  Their fluctuations form a tower of
fields, due to the excitations in the $y$ direction, and due to
those in the $m$-dimensional compact space. Their masses are
obtained. They are independent of $p$ and the lowest one is of
the order of, but larger than, the TeV scale. We also find that
for any bulk or brane field, {\em e.g.} graviton, $m$-form
field, or Higgs, the excitations in the compact space have 
Planckian mass and are likely to reintroduce the hierarchy
problem. 

Throughout in the following, we first present the results in a
form applicable to the general case, and then specialise to the
particular case under study. They are then applicable to the
case of a $D$ dimensional warped spacetime with bulk $m$-form
fields for arbitrary values of $D$ and $m$ and, thus, generalise
the analysis of \cite{gw2,gw0,hewett,pomarol}.

The plan of the paper is as follows. We present briefly the
relevent details of our scenario. We then analyse the radion
mass spectrum, fluctuation spectrum of the $m$-form field, and
their couplings to the matter fields living on a brane. We
conclude by mentioning a few open issues.

{\bf 2.}  
We consider $D = (n + m + 1)$ dimensional spacetime, $m \ge 1$,
containing flat $(n + m - 1)$ dimensional branes, with topology
${\bf R^{n - 1} \times T^m}$. We assume that ${\bf T^m}$ is of
$D$ dimensional Planckian size. Thus, on a macroscopic scale,
the branes are $(n - 1)$ dimensional, with $n = 4$ corresponding
to the observable case. 

In this letter, we study the Randall-Sundrum configuration
\cite{rs1} with the radion modulus stabilised by the component
of the $m$-form field along ${\bf T^m}$ \cite{k}. \footnote{In
our notation, $x^M = (x^\mu, y)$ denote the $D$ dimensional
spacetime coordinates, $x^\mu$, $\mu = 0, 1, \cdots, (D - 2)$,
the brane worldvolume coordinates, and $y$ the transverse
spatial coordinate. The signature of the metric is 
$(-, +, +, \cdots)$. The Riemann tensor is $R^M_{NKL} =
\partial_K \Gamma^M_{NL} + \cdots$. The $D$ dimensional Planck
mass $M_{pl}$ is set to unity, and the dimensionful quantities,
here and in the following, are all taken to be of ${\cal O}(1)$
unless mentioned otherwise.}  Thus, the transverse $y$ direction
is a circle, with $- y_1 \le y \le y_1$; the points $(x^\mu, y)$
and $(x^\mu, - y)$ are identified; and there are two branes
located at $y = 0$ and $y = y_1$, which are referred to in the
following as Planck and TeV branes respectively. The bulk fields
are the metric $g_{M N}$ and a totally antisymmetric $m$-form
field $B_{M_1 \cdots M_m}$, of mass $\Omega$.  The relevent
action is given by 
\begin{equation}\label{s}
S = \int d^D x \sqrt{- g} \left( \frac{R}{4} 
- \frac{G^2}{2 (m + 1)!} + V_0 
- \frac{\Omega^2 \chi}{2 m!} - \sum_{I = 0, 1} 
\delta(y - y_I) \Lambda_I(\chi) \right) \; , 
\end{equation} 
where $g = det(g_{MN})$, $G = d B$ is the field strength for
$B$, $V_0$ is a positive constant, and $\chi \equiv B_{M_1
\cdots M_m} B^{M_1 \cdots M_m}$. Also, $y_I$ are the locations
of the branes, and $\Lambda_I$ the brane potentials.

The background metric, ignoring the back reaction of $B$,  
is given by \cite{rs1} 
\begin{equation}\label{rs}
d s^2 = e^{2 A} \eta_{\mu \nu} 
d x^\mu d x^\nu + d y^2 
\end{equation} 
where $A = - k |y|$, 
$k^2 = \frac{4 V_0}{(D - 1)(D - 2)}$, and 
$\Lambda_0 = - \Lambda_1 = (D - 2) k$. 
The scale on the brane at $y = 0$ is taken to be the $D$
dimensional Planck scale. Then the scale on the brane at
$y = y_1$ is given by $e^{- k y_1}$, which is ${\cal O}(TeV)$ if
$k y_1 \simeq 40$.

As shown in \cite{k}, following the analysis of \cite{gw}, the
radion modulus $y_1$ can be stabilised at the required value
with no fine tuning by the component of the $m$-form field along
${\bf T^m}$. Briefly, the relevent details are as follows. 

Let the $m$-form field $B$ have non vanishing components 
only along ${\bf T^m}$, and the brane
potentials be such as to enforce the boundary conditions 
$\chi = \chi_0$ ($\chi_1$) at $y = 0$ ($y_1$). Furthermore, let 
$\frac{\chi_0}{m!}, \frac{\chi_1}{m!} \ll 1$ so that the back
reaction on the metric can be consistently neglected. 
The $m$-form field equation in the background metric 
(\ref{rs}) can then be solved. 
Substituting the solution into the 
action and integrating over the $y$ coordinate then yields an
effective potential $U$ for the modulus $y_1$. Defining 
\begin{equation}\label{defn} 
\gamma = \sqrt{\frac{\chi_1}{\chi_0}} 
\; , \; \; \; 
\frac{K}{k} = \frac{n + m}{2} + \tilde{\nu} 
\; , \; \; \; 
\epsilon = - \frac{n + m}{2} + \tilde{\nu} 
\end{equation}
where $\tilde{\nu} = \sqrt{\frac{(n - m)^2}{4} + 
\frac{\Omega^2}{k^2}}$, the potential $U(y_1)$ can be written as 
\begin{equation}\label{u}
U (y_1) = \frac{\chi_0 (u_1 + u_2)} 
{m! (1 - e^{- (K + \epsilon k) y_1})} 
\end{equation}
where $u_1 = e^{- (K - \epsilon k) y_1} (K - m k) (e^{-\epsilon
k y_1} - \gamma)^2$ and $u_2 = (m + \epsilon) k (1 - \gamma 
e^{- K y_1})^2$.  The radion modulus is stabilised at $y_1
\equiv y_{1s}$ where the potential $U(y_1)$ is minimum. The
required value for $y_{1s}$ can then be obtained with no fine
tuning (see \cite{k} and below).
 
{\bf 3.}  
We now analyse the radion fluctuations and determine its mass. 
A rigorous analysis of various fluctuations is given in
\cite{tanaka,csaki} which, however, requires an exact solution
that includes the back reaction of the radion stabilising field
on the metric. Such solutions are known in the case of the bulk
scalar field, but not in the present case.  Hence, we analyse
the radion fluctuations perturbatively as in \cite{gw2}, 
neglecting the back reaction on the metric.

The radion field $T(x)$ modifies the metric (\ref{rs}) as
follows: 
\begin{equation}\label{radionmetric}
d s^2 = e^{- 2 k |y| T(x)} g_{\mu \nu}(x)  
d x^\mu d x^\nu + T^2(x)d y^2 \; . 
\end{equation} 
One next calculates the Ricci scalar $R$ and reduces the gravity
action from $D$ to $(D - 1)$ dimensions. For this purpose,
consider a general metric of the form 
\[
d s^2 = e^{P(x,y)} g_{\mu \nu}(x) d x^\mu d x^\nu 
+ e^{Q(x,y)} h_{a b}(y) d y^a d y^b 
\] 
where $\mu, \nu = 0, 1, \cdots, (d_1 - 1)$ and 
$a, b = 1, 2, \cdots,d_2$. After some algebra, 
the Ricci scalar $R$ for the above metric
can be written compactly as 
\begin{equation}\label{d1d2}
R = R_{\tilde{g}} - d_2 \nabla^2_{\tilde{g}} Q 
- \frac{d_2 (d_2 + 1)}{4} e^{- P} (\nabla_g Q)^2 
+ \; (d_1, P, g) \longleftrightarrow 
(d_2, Q, h)  
\end{equation}  
where the last part on the right hand side of (\ref{d1d2}) is
obtained from the first by interchanging $d_1$ and $d_2$, $P$
and $Q$, and $g_{\mu \nu}$ and $h_{a b}$. $R_{\tilde{g}}$ and
$\nabla^2_{\tilde{g}} Q$ are the Ricci scalar and the Laplacian
on $Q$ for the conformally transformed metric
$\tilde{g}_{\mu \nu} = e^P g_{\mu \nu}$. Explicitly, 
\begin{eqnarray*}
R_{\tilde{g}} & = & e^{- P} ( R_g - (d_1 - 1) \nabla_g^2 P 
- \frac{(d_1 - 1) (d_1 - 2)}{4} (\nabla_g P)^2 ) \\ 
\nabla^2_{\tilde{g}} Q & = & e^{- P} ( \nabla_g^2 Q 
+ \frac{(d_1 - 2)}{2} (\nabla_g P) (\nabla_g Q) ) 
\end{eqnarray*}
where the subscript $g$ denotes that the corresponding
quantity is to be calculated using the metric $g_{\mu \nu}$. 
Similarly for $R_{\tilde{h}}$ and $\nabla^2_{\tilde{h}} P$.  

The Ricci scalar $R$ can now be calculated for the metric
(\ref{radionmetric}) where $d_1 = D - 1$, $d_2 = 1$,   
$P = - 2 k |y| T(x)$, $Q = 2 ln T(x)$, and $h_{1 1} = 1$. One
then substitutes this expression in the gravity part of the
action, and integrates over the $y$ coordinate. Carrying out
these steps, which are straightforward, yields the following 
$(D - 1)$ dimensional action for the radion field $T$: 
\[
S =  \int d^{D - 1} x \sqrt{- g} \left(
\frac{(1 - e^{- (D - 3)k y_{1s} T}) R_g}{2 k (D - 3)} 
- \frac{2 (D - 2)(\nabla_g 
e^{- \frac{D - 3}{2} k y_{1s} T})^2}
{k (D - 3)^2}  \right) 
\] 
where $y_{1s}$ is the stabilised value of the modulus. However,
$T$ is not canonically normalised. The canonically normalised
radion field, denoted by $\phi$, is instead given by 
\begin{equation}\label{radion} 
\phi = f e^{- \frac{D - 3}{2} k y_{1s} T} \; , \; \; \; 
f \equiv \sqrt{\frac{4 (D - 2)}{k (D - 3)^2}} \; . 
\end{equation}
The resulting $(D - 1)$ dimensional action for the field $\phi$
then becomes 
\begin{equation}\label{sradion}
S = \int d^{D - 1} x \sqrt{- g} \left( 
\frac{(1 - \frac{\phi^2}{f^2}) R_g}{2 k (D - 3)} 
- \frac{(\nabla_g \phi)^2}{2} \right) \; . 
\end{equation}

The stabilisation mechanism for the radion modulus induces a
potential $U$ for the radion field which has a minimum at 
$y_1 \equiv y_{1s}$ corresponding to $<T>_{VEV} = 1$. 
The radion mass is then given by
\begin{equation}\label{*}
m^2_{radion} = 
\frac{\partial^2 U}{\partial \phi^2}(y_{1s}) \; .  
\end{equation} 

The above expressions are general and are valid for any value of
$D$, irrespective of the stabilisation mechanism. For example,
they reduce to those given in \cite{gw2} for $D = 5$. For
the case considered here, $D = (n + m + 1)$ and the radion
potential $U$ is given by equation (\ref{u}) \cite{k}.
$m^2_{radion}$ can then be obtained from (\ref{*}).

However, obtaining exact analytic expressions for $y_{1s}$ and
$m^2_{radion}$ for the potential $U$ given in (\ref{u}) is
difficult. Hence, we proceed as follows. (The algebraic steps
involved in obtaining equations (\ref{ucan})-(\ref{mradion})
below are straightforward and, hence, are omitted.) Rearranging
the terms suitably, we first rewrite $U$ as 
\begin{equation}\label{ucan}
U = \frac{(m + \epsilon) k \chi_0}{m!} 
+ \frac{\chi_0 e^{- (K - \epsilon k) y_1}}{m!} \; 
\left( \frac{(K + \epsilon k) 
(e^{- \epsilon k y_1} - \gamma)^2}
{(1 - e^{- (K + \epsilon k) y_1})} 
- (m + \epsilon) k \gamma^2 \right) \; . 
\end{equation} 
The analysis of \cite{gw,gw2,k} then essentially amounts to
neglecting the last term above.  However, a much better
approximation is to neglect instead the exponential factor in
the denominator. It is then straightforward to obtain $y_{1s}$
and $m^2_{radion}$. Neglecting this factor and setting $\frac{d
U}{d y_1}$ to zero then gives 
\[
(K + \epsilon k)^2 e^{- 2 \epsilon k y_1} 
- 2 \gamma K (K + \epsilon k) e^{- \epsilon k y_1} 
+ \gamma^2 (K - \epsilon k) (K - m k) = 0 
\]
which is just a quadratic equation for $e^{- \epsilon k y_1}$.
The solution $y_1 \equiv y_{1s}$ where $U$ is minimum is then
given by
\begin{equation}\label{y1scan}
k y_{1s} = - \frac{1}{\epsilon} \; 
ln \left( \frac{\gamma (n + m + \epsilon + c)}
{n + m + 2 \epsilon} \right) 
\; , \; \; \; \; 
c = \sqrt{\epsilon^2 + (n + m) (m + \epsilon)} 
\end{equation} 
where we have used $K = (n + m + \epsilon) k$ which follows
from equation (\ref{defn}). From (\ref{y1scan}), it can be seen
that $k y_{1s} \simeq 40$ can be easily achieved without any
fine tuning by choosing $\epsilon$ to be small \cite{gw,k}. In
the following we assume that $\epsilon$ is small and that the
parameters $\gamma$ and $\epsilon$ are chosen such that $k
y_{1s} \simeq 40$.

One can now evaluate the second derivative of $U$ at $y_{1s}$
and, thereby, obtain the radion mass.  Using equations
(\ref{*}), (\ref{ucan}), and (\ref{y1scan}), it is given by 
\begin{equation}\label{mradion} 
m^2_{radion} = \frac{2 (n + m + \epsilon + c)}{(n + m - 1)} \; 
\left( \frac{\chi_1 \epsilon c}{m!} \right) \; 
k^2 e^{- 2 k y_{1s}} 
\end{equation}
which gives the radion mass in the present scenario where 
the radion is stabilised by a bulk $m$-form field. 

Consider equation (\ref{mradion}). The exponential factor brings
the radion mass down to the TeV scale, namely to the scale 
$k e^{- k y_{1s}}$ on the TeV brane. It is further suppressed by
$\frac{\chi_1}{m!}$, which is taken to be $\ll 1$ so that the
above analysis is valid where the back reaction is neglected.
Hence, the radion mass will be smaller than TeV. These
suppression factors are also present in the scenario where
radion is stabilised by the bulk scalar field
\cite{randall,gw2}.

Consider now the factor $(\epsilon c)$. This suppression of the
radion mass by the $\epsilon$ factor is analogous to that found
in the case of the bulk scalar field \cite{randall,gw2}.
However, the exponent of $\epsilon$ in these cases are
different.

The results for the bulk scalar field case can be obtained from
the above expressions by setting $m = 0$ formally. Then, with 
$m = 0$ and $\epsilon$ small, $c \simeq \sqrt{\epsilon}$ and, 
as found in \cite{randall} and mentioned in the footnote 2 of
\cite{gw2}, 
\begin{equation}\label{3/2}
m^2_{radion} \propto \epsilon c \simeq 
\epsilon^{\frac{3}{2}} \; . 
\end{equation} 
The above expressions are also valid when the last term in
(\ref{ucan}) is neglected if the $(m + \epsilon)$ factor in the
expression for $c$ is formally set to zero. Then, 
$c = \epsilon$ for any value of $m$, and we obtain the result
given in \cite{gw2}:
\begin{equation}\label{2}
k y_{1s} = - \frac{1}{\epsilon} ln \gamma 
\; , \; \; \; {\rm and} \; \; \; 
m^2_{radion} \propto \epsilon^2 \; . 
\end{equation}

Note that the exact analysis of \cite{tanaka,csaki} also gives
the above relation for the radion mass, but the origin of the
different scalings in (\ref{3/2}) and (\ref{2}) is not
understood \cite{csaki}.  For the case of interest here, 
$m \ne 0$ and $c \simeq \sqrt{m (n + m)} = {\cal O}(1)$. Then 
\begin{equation}\label{1}
m^2_{radion} \propto \epsilon \; .
\end{equation}

Thus, in the perturbative analysis of the present scenario, the
radion mass is given by equation (\ref{mradion}). It is smaller
than TeV but larger, by a factor of $\epsilon^{- \frac{1}{4}}$
($\epsilon^{- \frac{1}{2}}$), than that obtained in
\cite{randall,gw2} (\cite{gw2,tanaka,csaki}) where the radion is
stabilised by a bulk scalar field. 

{\bf 4.} 
We now consider the fluctuations of the other components of the
$m$-form field. If the $m$-form field has non vanishing
components along $p$ of the ${\bf R^n}$ directions, $1 \le p \le
min (n, m)$, then, from the macroscopic $n$ dimensional point of
view, it mimics a $p$-form field. The number of such $p$-form
fields is equal to $\frac{m!}{p! (m - p)!}$. Here, we analyse
the fluctuation spectrum of these fields.

Consider the general case of an $m$-form bulk field $B_{M_1
\cdots M_m}$, of mass $M$, in the $D$ dimensional spacetime with
the background warped metric given by (\ref{rs}). The action
for the $m$-form field $B$ is given by  
\begin{equation}\label{sp} 
S = - \int d^{D - 1} x d y \sqrt{- g} 
\left( \frac{G^2}{2 (m + 1)!} 
+ \frac{\Omega^2 \chi}{2 m!} \right) 
\end{equation} 
where $G$ and $\chi$ are as defined in equation (\ref{s}). The
equation of motion is given by 
\begin{equation}\label{eomp}
\nabla_M G^{M M_1 \cdots M_m} = M^2 B^{M_1 \cdots M_m} \; . 
\end{equation} 
For the background metric (\ref{rs}), the action (\ref{sp}) can
be written explicitly as 
\begin{eqnarray}
S & = & - \int d^{D - 1} x d y \; e^{(D - 1 - 2 m) A} 
\nonumber \\ 
& & \times \left(
\frac{e^{- 2 A} (G_{\mu \mu_1 \cdots \mu_m})^2}
{2 (m + 1)!}  
+ \frac{(\partial_y B_{\mu_1 \cdots \mu_m})^2 
+ \Omega^2 (B_{\mu_1 \cdots \mu_m})^2}{2 m!} \right) \; , 
\label{sp'} 
\end{eqnarray}
where the $\mu$ indices are now contracted using 
$\eta^{\mu \nu}$. Consider the ansatz  
\begin{equation}\label{bxy}
B_{\mu_1 \cdots \mu_m} (x^\mu, y) = 
\sum_i B^{(i)}_{\mu_1 \cdots \mu_m} (x^\mu) 
\frac{f_{(i)}(y)}{\sqrt{y_1}}   
\end{equation}
where the functions $f_{(i)}(y)$ 
satisfy the differential equation 
\begin{equation}\label{feqn'}
- \frac{d}{d y} (e^{(D - 1 - 2 m) A} 
\frac{d}{d y} f_{(i)}) + M^2 e^{(D - 1 - 2 m) A} 
f_{(i)} = m^2_{(i)} e^{(D - 3 - 2 m) A} f_{(i)}  
\end{equation}
and are normalised as follows: 
\begin{equation}\label{norm}
\int_{- y_1}^{y_1}  \frac{dy}{y_1} e^{(D - 3 - 2 m) A} 
f_{(i)}(y) f_{(j)}(y) = \delta_{i j} \; . 
\end{equation} 
Since the points $(x^\mu, y)$ and $(x^\mu, - y)$ are identified,
it follows that $f_{(i)}(y)$ must satisfy the conditions
$f_{(i)}(- y) = f_{(i)}(y)$, which determines $f_{(i)}(- y)$,
and 
\begin{equation}\label{bc}
\frac{d f_{(i)}}{d y}(0) 
= \frac{d f_{(i)}}{d y}(y_1 ) = 0 \; . 
\end{equation} 
Therefore, it suffices to obtain $f_{(i)}(y)$ for 
$0 \le y \le y_1$ satisfying the boundary condition (\ref{bc}).

Using equations (\ref{bxy}), (\ref{feqn'}), and (\ref{norm}),
the $(D - 1)$ dimensional action for the fields $B^{(i)}(x^\mu)$
can now be determined.  Partial integrating the $(\partial_y
B)^2$ term in (\ref{sp'}), and then performing the $y$
integration using the above equations, the $(D - 1)$ dimensional
action for the fields $B^{(i)}$ becomes 
\begin{equation}\label{spd-1}
S = - \int d^{D - 1} x \sum_i \left(
\frac{(G^{(i)})^2}{2 (m + 1)!} 
+ \frac{m^2_{(i)} (B^{(i)})^2}{2 m!} \right) \; 
\end{equation} 
which shows that $m_{(i)}$ is the mass of the $(D - 1)$
dimensional $m$-form fields $B^{(i)}$. (Substituting the ansatz
(\ref{bxy}) directly into the equation of motion (\ref{eomp})
also leads to the equation (\ref{feqn'}) for $f_{(i)}(y)$, and
to the equation of motion for $B^{(i)}$ obtained from
(\ref{spd-1}).)

Consider equation (\ref{feqn'}) for $y \ge 0$. Using 
$A = - k y$, its solutions are given in terms of Bessel
functions $J$ and $Y$ by 
\begin{equation}\label{fsoln}
f_{(i)} = \frac{e^{\frac{D - 1 - 2 m}{2} k y}}{N_{(i)}} 
\; \left( J_\nu(\frac{m_{(i)}}{k} e^{k y}) 
+ b_{(i)} Y_\nu(\frac{m_{(i)}}{k} e^{k y}) 
\right) \; , 
\end{equation}
where $b_{(i)}$ and $N_{(i)}$ are constants and $\nu =
\sqrt{\frac {(D - 1 - 2 m)^2}{4} + \frac{M^2}{k^2}}$. It
follows from equation (\ref{norm}) that the normalisation 
constant $N_{(i)}$ is given by  
\begin{equation}\label{normsoln}
N^2_{(i)} = \int_{- y_1}^{y_1} \frac{dy}{y_1} 
e^{2 k y } \left( J_\nu(\frac{m_{(i)}}{k} e^{k y}) 
+ b_{(i)} Y_\nu(\frac{m_{(i)}}{k} e^{k y}) 
\right)^2 \; . 
\end{equation}
Using (\ref{fsoln}), the boundary condition (\ref{bc}) 
implies that 
\begin{equation}\label{bcb}
b_{(i)} = b(\frac{m_{(i)}}{k}) 
= b(\frac{m_{(i)}}{k} e^{k y_1})
\end{equation}
where we have defined $b(x) \equiv - \frac {(D - 1 - 2 m)
J_\nu(x) + 2 x J'_\nu(x)} {(D - 1 - 2 m) Y_\nu(x) + 2 x
Y'_\nu(x)}$, with the primes denoting the derivatives with
respect to $x$. These relations determine the constant $b_{(i)}$
and the eigenvalues $m_{(i)}$.

The eigenvalues of the masses $m_{(i)}$ can be estimated easily
from the above equations. Note that when the radion modulus is
stabilised, as in the present case, $k y_1 = k y_{1 s} \simeq
40$. The exponent $e^{k y_1}$ is therefore extremely large. It
then follows that $m_{(i)}$ and $b_{(i)}$ are extremely small,
and that the mass eigenvalues are given approximately by 
\begin{equation}\label{mi}
m_{(i)} \simeq x_{\nu i} \; (k e^{- k y_{1s}}) 
\simeq x_{\nu i} \; \times \; {\rm TeV}  
\end{equation}  
where $x_{\nu i}$ is the $i^{th}$ zero of the Bessel function
$J_\nu$. Hence, it follows that, in the background of the
warped metric (\ref{rs}), the fluctuations of the $m$-form field
has a mass spectrum given by (\ref{mi}). The normalisation
constant $N_{(i)}$ can also be estimated from (\ref{normsoln}).
Since $b_{(i)}$ is small and $k y_{1s} \gg 1$, $N_{(i)}$ is
given approximately by \cite{gw0,pomarol} 
\begin{equation}\label{normapp}
N^2_{(i)} \simeq  \int \frac{dy}{y_1} e^{2 k y } 
J^2_\nu(\frac{m_{(i)}}{k} e^{k y}) 
\simeq \frac{e^{2 k y_{1s}}}{k y_{1s}} \; . 
\end{equation}
The above expressions are general and are valid for any value of
$D$, $m$, and $M$. For example, they reduce to those given in
\cite{gw0} (\cite{hewett,pomarol}) for $D = 5$ and $m= 0$ ($1$),
which corresponds to a bulk scalar (vector) field. 

In the present case $D = (n + m + 1)$ and $M^2 = \Omega^2$.
Hence, as follows from equation (\ref{defn}), the index 
$\nu = \tilde{\nu} > \frac{n + m}{2}$. Since the first zero
$x_{\nu 1} > \nu$ \cite{watson}, the lowest mass 
$m_{(1)} > \frac{n + m}{2} \; (k e^{- k y_{1s}}) 
\stackrel{>}{_\sim} TeV$. 

In the present case, the brane is $(n + m - 1)$ dimensional,
with topology ${\bf R^{n - 1} \times T^m}$, where ${\bf T^m}$ is
of Planckian size. Hence, from the macroscopic $n$ dimensional
spacetime point of view, the Kaluza-Klein (KK) excitations of
both the bulk and the brane fields along ${\bf T^m}$ will form a
tower of fields. Consider the bulk $B$ field. (Other fields such
as graviton, Higgs, etc. can also be analysed similarly, and the
results are qualitatively the same.)  Let $\tilde{x}$ be the
coordinates of the $n$ dimensional spacetime, $\xi^a$ and $L_a$
be, respectively, the coordinate and the size of the $a^{th}$
direction of ${\bf T^m}$, and $n_a$ the corresponding KK
excitation number. The KK excitations can then be written
schematically as 
\begin{equation}\label{kkansatz} 
B^{(i)}(x^\mu) = \sum_{(n_a)} B^{(n_a,i)}(\tilde{x}) 
e^{i \sum_a \frac{n_a \xi^a}{L_a}} \; . 
\end{equation}
Substituting (\ref{kkansatz}) and (\ref{bxy}) in the action
(\ref{sp}) and integrating over the $\xi^a$ and $y$ coordinates,
in any order, one obtains the $n$ dimensional action for the
tower of fields $B^{(n_a,i)}$, given by
\begin{equation}\label{spn}
S = - \int d^n \tilde{x} \sum_{n_a,i} 
\left( \frac{(G^{(n_a,i)})^2}{2 (m + 1)!} 
+ \frac{m^2_{(n_a,i)} (B^{(n_a,i)})^2}{2 m!} \right) 
\end{equation}
where the masses $m_{(n_a,i)}$ of the fields $B^{(n_a,i)}$
are given by 
\begin{equation}\label{mnai}
m^2_{(n_a,i)} = m^2_{(i)} + M_{KK}^2 \; , \; \; \; 
M_{KK}^2 \equiv \sum_a \frac{n_a^2}{L_a^2} \; . 
\end{equation}
(Substituting the ansatz (\ref{kkansatz}) and (\ref{bxy})
directly into the equation of motion (\ref{eomp}) also leads to
the above result.)  Note that $m_i$ is of order $TeV$, as given
by (\ref{mi}), whereas $M_{KK}$ is of order $M_{pl}$ if $n_a$
are not all zero.

The above formula applies to all the components of the $m$-form
field which, from the macroscopic $n$ dimensional spacetime
point of view, mimics the $\frac{m!}{p! (m - p)!}$ number of
$p$-form fields, $1 \le p \le min(n, m)$. Thus, it follows that
their fluctuations form a tower of fields, with masses
independent of $p$ and given by (\ref{mi}) and (\ref{mnai}).
The zero modes along ${\bf T^m}$ have masses given by (\ref{mi})
and are of the order of, but larger than, $TeV$. The KK
excitations have masses given by (\ref{mnai}) and are of order
$M_{pl}$. 

The fact that $KK$ excitations along ${\bf T^m}$ have Planckian
mass can also be seen as follows. Because of the warped metric
(\ref{rs}), the size of ${\bf T^m}$ at $y$ is $e^{- k y}$ times
the Planck length. Hence, the KK excitations will have a mass of
order $e^{k y} M_{pl}$\footnote{We thank the referee for the
above comments in this paragraph.}. However, their kinetic
terms are not in the canonical form and, therefore, the fields
need to be rescaled. Upon this rescaling, the masses are
suppressed by a factor of $e^{- k y}$ \cite{rs1}.  The KK
excitation mass will then be of order $M_{pl}$. This can be seen
explicitly if one substitutes (\ref{kkansatz}) in (\ref{sp}) and
integrates over the $\xi^a$ coordinates first.

This phenomenon is generic for any bulk or brane field if the
braneworld has compact directions of Planckian size. An
important consequence is that it is very likely to reintroduce
the hierarchy problem.  For example, both graviton and the Higgs
field will have KK excitations, with mass of order
$M_{pl}$. From the macroscopic $n$ dimensional spacetime point
of view, the zero mode of the Higgs field can couple directly to
its own KK excitations through the Higgs potential, and also to
the graviton KK excitations through their mutual gravitational
couplings. Even if the corresponding coupling strengths are
suppressed by Planck scale, the loop corrections are still
likely to push up the Higgs mass to $M_{pl}$ because of the
infinite number of fields present in the KK tower, thereby
reintroducing the hierarchy problem. Explicit calculations are
currently in progress. 

To avoid this problem, one may wish to take the size of ${\bf
T^m}$ to be $\stackrel{>}{_\sim} k^{- 1} e^{k y_{1s}} \simeq
(TeV)^{- 1}$, so that it is always larger than the Planck length
for $- y_{1s} \le y \le y_{1s}$.  KK excitation mass would then
be of order $k e^{- k y_{1s}} \simeq TeV$, and the lowest
excitation, {\em e.g.} of graviton, may then provide another
signature of the model. However, stabilising the size of ${\bf
T^m}$ at such a value introduces a different hierarchy problem
since the fundamental scale is assumed to be
$M_{pl}$\footnote{We thank the referee for the above comments in
this paragraph.}.  Moreover, the $n$ dimensional Newton's
constant $G_N$, given by $G_N^{2 - n} \propto Vol({\bf T^m})
k^{- 1} M_{pl}^{n + m - 1}$, will not be 
${\cal O}(M_{pl}^{- 1})$. For these reasons, we assume that
${\bf T^m}$ is of Planckian size.

{\bf 5.} 
Consider the couplings to matter fields on the brane.  The
radion couplings in the case of bulk scalar field have been
analysed in detail in \cite{randall,gw2,tanaka,csaki}. The same
analysis applies for the radion couplings in the present case
also. Hence, we will not consider it further.

We analyse here the couplings of the other components of
the $m$-form fields to the matter fields living on the brane
located at $y = y_*$. Let ${\cal O}^{M_1 \cdots M_m}$, $M_1,
M_2, \cdots = 0, 1, \cdots, (D - 2)$ be the matter field
operator on the brane at $y_*$ to which the $m$-form field
couples \footnote{For example, for fermionic coupling, 
${\cal O}^{M_1 \cdots M_m} \sim \bar{\Psi} \Gamma^{M_1} \cdots
\Gamma^{M_m} \Psi$ where $\Psi$ denotes the fermion living on
the brane and $\Gamma$ the Dirac matrix.}. The relevent action
is given by 
\begin{equation}\label{scoup}
S_{{\cal O}} = \int d^{D - 1} x d y \sqrt{- g} \; 
(\lambda_D {\cal O}^{M_1 \cdots M_m}) \; 
B_{M_1 \cdots M_m} 
\delta(y - y_*) \; , 
\end{equation} 
where $\lambda_D$ is the coupling constant. 

As will be seen below, this action leads 
to the $(D - 1)$ dimensional
action of the form 
$S_{{\cal O}} \sim \int d^{D - 1} x \; 
(\lambda {\cal O}^{\mu_1 \cdots \mu_m}) \; 
\sum_i B^{(i)}_{\mu_1 \cdots \mu_m}$ where the indices are
contracted using $\eta^{\mu \nu}$. With no loss of
generality, we define the coupling constant $\lambda$ to be
dimensionless, which can always be done by multiplying the
operator ${\cal O}$ with suitable powers of the Planck mass
$M_{pl}$, which has been set to unity in our notation here. 
The mass dimensions, denoted by ${[} \;\; {]}$, 
of the various quantities can now be obtained: 
${[} \lambda {]} = 0$ by assumption, and 
${[} B^{(i)} {]} = \frac{D - 3}{2}$ as follows from 
(\ref{spd-1}). Therefore, 
${[} {\cal O} {]} = \frac{D + 1}{2}$. \footnote{Thus, 
for example, since ${[} \Psi {]} = \frac{D - 2}{2}$ 
the fermionic operator is given by 
${\cal O}^{M_1 \cdots M_m} = M^{\frac{5 - D}{2}}_{pl} 
\bar{\Psi} \Gamma^{M_1} \cdots \Gamma^{M_m} \Psi$, 
where the Planck mass dependence is chosen so that 
${[} {\cal O} {]} = \frac{D + 1}{2}$.}   
Also, as can be seen from equation (\ref{bxy}), 
${[} B {]} = \frac{D - 2}{2}$ and, hence, 
${[} \lambda_D {]} = - \frac{1}{2}$.

Consider now the action (\ref{scoup}). Writing ${\cal O}^{M_1
\cdots M_m} = E^{M_1}_{a_1} \cdots E^{M_m}_{a_m} {\cal O}^{a_1
\cdots a_m}$, where $E^M_a \propto e^{k y_*}$ are the $D$-beins,
and using equations (\ref{bxy}) and (\ref{fsoln}), the action
(\ref{scoup}) can be written as 
\begin{eqnarray*}
S_{{\cal O}} & = & \int d^{D - 1} x \; 
e^{- \frac{D - 1}{2} k y_*} \; (\lambda 
{\cal O}^{\mu_1 \cdots \mu_m}) \\
& & \times \sum_i 
\frac{B^{(i)}_{\mu_1 \cdots \mu_m}}{N_{(i)}} 
\left( J_\nu(\frac{m_{(i)}}{k} e^{k y_*}) 
+ b_{(i)} Y_\nu(\frac{m_{(i)}}{k} e^{k y_*}) \right) 
\end{eqnarray*}
where $\lambda \equiv \frac{\lambda_D}{\sqrt{y_{1s}}}$. 
Because of the warped nature of the $D$ dimensional spacetime,
the scale on the brane at $y_*$ is set by $e^{- k y_*} M_{pl}$.
Since ${[} {\cal O} {]} = \frac{D + 1}{2}$, the appropriate
operator ${\cal O}_*$, with all its mass scales corresponding to
the scale $e^{- k y_*}$ on the brane at $y_*$, is given by
\[ 
{\cal O}_*^{\mu_1 \cdots \mu_m} = 
e^{- \frac{D + 1}{2} k y_*} \; 
{\cal O}^{\mu_1 \cdots \mu_m} \; .  
\] 
Using the fact that $b_{(i)}$ is negligible and using
(\ref{normapp}) for $N_{(i)}$, we then obtain
\begin{equation}\label{so*} 
S_{{\cal O}_*} \simeq \int d^{D - 1} x 
e^{- k (y_{1s} - y_*)} \; (\lambda \sqrt{k y_{1s}}) \; 
{\cal O}_*^{\mu_1 \cdots \mu_m} \; 
\sum_i B^{(i)}_{\mu_1 \cdots \mu_m} 
J_\nu(\frac{m_{(i)}}{k} e^{k y_*})   
\end{equation} 
which describes the coupling of the $(D - 1)$ dimensional
$m$-form fields $B^{(i)}$ to the matter fields living on the
brane at $y_*$.

From equation (\ref{so*}), the following features can be 
seen easily:
{\bf (i)} 
The dimensionless coupling constant $\lambda$ is enhanced by a
factor $\sqrt{k y_{1s}}$, which arises from the normalisation
constant $N_{(i)}$.
{\bf (ii)}
Near the Planck brane, $y_* \simeq 0$ and the coupling is
suppressed by $e^{- k y_{1s}}$, which also arises from 
$N_{(i)}$. It is further suppressed by the factor 
$J_\nu \simeq \left(\frac{m_{(i)}}{k}\right)^\nu \simeq 
e^{- \nu k y_{1s}}$. 
{\bf (iii)} 
Near the TeV brane, $y_* \simeq y_{1s}$ and, hence, both the
suppression factors described in {\bf (ii)} are absent. Only the
enhancement factor $\sqrt{k y_{1s}}$ is present.

The above expressions are general and are valid for any value of
$D$ and $m$. It is easy to check that they reduce to those given
in \cite{hewett,pomarol} for $D = 5$ and $m= 1$, which
corresponds to a bulk vector field. In the present case, 
$D = (n + m + 1)$ and the action (\ref{so*}) describes the
couplings of the $m$-form field to the matter fields living on
the brane located at $y_*$.

{\bf 6.}  
We now conclude by mentioning a few open issues. The radion mass
is obtained by a perturbative analysis. A rigorous analysis
requires an exact solution including the back reaction of the
$m$-form field on the metric which, however, is not known at
present.  It is thus important to find such solutions, which may
be of interest in their own right. Also, it will be of interest
to study other phenomenological features which can distinguish
the present scenario from that of \cite{rs1,gw}.

The KK excitations of Planckian mass are generically present if
the braneworld has compact directions of Planckian size. They
are very likely to reintroduce the hierarchy problem. Explicit
calculations of this phenomenon within the present scenario
are currently in progress.

However, since Planckian lengths are involved, a consistent
study and possible resolution, if any, of this problem requires
the knowledge of the underlying theory.  It is, therefore,
important to answer the question of whether the present scenario
can be realised in a fundamental theory, {\em e.g.}
supergravity, string, or M theory. Although the answer is
presently not known, it is encouraging that massive $m$-form
fields appear naturally in massive type IIA supergravity
theories \cite{romans,howe}, which are believed to be related to
M theory. It is then worthwhile to study whether the present
scenario can be realised by some suitable compactification of
these theories.

\vspace{3ex}

{\bf Acknowledgement:} 
It is a pleasure to thank R. K. Kaul for discussions, and also
the referee for his/her important and helpful comments.


\end{document}